\documentclass[preprints]{sigchi}

\CopyrightYear{2020}
\doi{none}
\isbn{none}

\usepackage{balance}       
\usepackage{graphics}      
\usepackage[T1]{fontenc}   
\usepackage{txfonts}
\usepackage{mathptmx}
\usepackage[pdflang={en-US},pdftex]{hyperref}
\usepackage{color}
\usepackage{booktabs}
\usepackage{textcomp}
\usepackage{enumitem}
\usepackage{tabularx}

\usepackage{microtype}        
\usepackage{ccicons}          

\usepackage{todonotes}

\def\plaintitle{Let's Gamble: Uncovering the Impact of Visualization on Risk Perception and Decision-Making}

\def\emptyauthor{}
\def\plainkeywords{Visualization; Decision-Making; Proportion Estimates; Risk Behavior; Decision Theory}

\makeatletter
\def\url@leostyle{%
  \@ifundefined{selectfont}{
    \def\UrlFont{\sf}
  }{
    \def\UrlFont{\small\bf\ttfamily}
  }}
\makeatother
\def\pprw{8.5in}
\def\pprh{11in}

\definecolor{linkColor}{RGB}{6,125,233}
\hypersetup{%
  pdftitle={\plaintitle},
  pdfauthor={\emptyauthor},
  pdfkeywords={\plainkeywords},
  pdfdisplaydoctitle=true, 
  bookmarksnumbered,
  pdfstartview={FitH},
  colorlinks,
  citecolor=black,
  filecolor=black,
  linkcolor=black,
  urlcolor=linkColor,
  breaklinks=true,
  hypertexnames=false
}

\begin{document}

\title{\plaintitle}

\numberofauthors{3}
\author{%
  \alignauthor{Melanie Bancilhon\\
    \affaddr{mbancilhon@wustl.edu}\\
    \affaddr{\hspace{2mm}}\\
    \email{\hspace{2mm}}}\\
  \alignauthor{Zhengliang Liu\\
    \affaddr{zhengliang@wustl.edu}\\
    \affaddr{\hspace{2mm}}\\
    \email{Washington Univ.\ in St.\ Louis}}\\
  \alignauthor{Alvitta Ottley\\
    \affaddr{alvitta@wustl.edu}\\
    \affaddr{\hspace{2mm}}\\
    \email{\hspace{2mm}}}\\
}
\maketitle

\begin{abstract}

Data visualizations are standard tools for assessing and communicating risks. However, it is not always clear which designs are optimal or how encoding choices might influence risk perception and decision-making. In this paper, we report the findings of a large-scale gambling game that immersed participants in an environment where their actions impacted their bonuses. Participants chose to either enter a draw or receive guaranteed monetary gains based on five common visualization designs. By measuring risk perception and observing decision-making, we showed that icon arrays tended to elicit economically sound behavior. We also found that people were more likely to gamble when presented area proportioned triangle and circle designs. Using our results, we model risk perception and decisions for each visualization and provide a ranking to improve visualization selection.

\end{abstract}




\section{Introduction}
\label{sec:introduction}
There are several competing views on what it means to make a decision.
Psychologists believe that we make choices based on empirical evidence and beliefs about the likelihood of specific events~\cite{tversky1992}. 
Economic theorists view decision-making as a selection between alternatives based on a weighted sum of probabilities~\cite{prelec1991}.  
Such choices include deciding whether to bike or drive based on the chance of rain, whether to opt for preventive health care based on the likelihood of developing a disease or whether to enter a gamble based on the chance of winning a prize. 


In many cases, it is increasingly common to use data visualization to support reasoning about risks and to aid sound decision-making.
However, it is reasonable to assume that encoding choices can influence the decisions people make.
Researchers have shown that visualizations can impact speed~\cite{cleveland1984graphical,simkin1987information}, accuracy~\cite{simkin1987information}, memorability~\cite{bateman2010useful,borkin2016beyond}, statistical reasoning~\cite{micallef2012assessing,ottley2016improving}, and judgement~\cite{cleveland1984graphical,correll2014error,zacks1998reading}. 
These effects can have significant repercussions, especially when decisions are life-altering. Still, a designer can represent the same data using different, yet equally theoretically valid visualization designs~\cite{mackinlay2007show}, and it is sometimes difficult for designers to identify a poor fit for the data~\cite{pandey2015deceptive}.
Therefore, what we need is a thorough understanding of which designs lead to (in)accurate judgment and how probability distortions influence behavior. 


At the heart of decision-making with visualization is graphical perception. People must first decode the information presented in graphical form in order to reason and make a choice. Early work has investigated the relationship between design and graphical perception~\cite{cleveland1984graphical,cleveland1987graphical} when making proportion judgment. We know, for example, that people are best at decoding quantitative information when the data is encoded in the position of a visual mark and worst when the encoding uses area. Subsequent studies that looked explicitly at part-of-the-whole judgment tasks found no difference between angle (pie charts) and position (bar graphs)~\cite{lewandowsky1989discriminating,simkin1987information,skau2016arcs}. In this paper, we reexamine basic encoding options to investigate how data visualization design might influence decision-making.


The work in this paper builds on the previous work in two ways. First, we investigate proportion judgment for five common chart designs across a range of seven probability values. By varying both the chart types and the probability values, we are able to provide guidance for how practitioners might choose a visual representation based on their data. Second, we utilize a gambling game that immersed participants in an environment where their actions impacted their bonuses. The lottery scenarios allow us to observe decisions with real-world consequences. Consider the following hypothetical gamble:

\begin{quote}\label{example-prospect}
\centering
\textit{Which do you prefer?}\newline\newline
\textit{A: 50\% chance to win \$1000, 50\% chance to win nothing}\newline
\textit{or}\newline
\textit{B: \$450 for sure}~\cite{kahneman2013prospect}\newline
\end{quote}
\noindent 

Decision theorists have long studied simple gambles such as this because gambles provide a straightforward framework that shares key characteristics with complex real-world situations~\cite{bruhin2010risk}. For instance, complex decisions are often a choice between one or more prospects, and the consequences of the decisions are almost always uncertain. The lottery scenario is a useful test-bed because one need only to weigh the risk of entering the lottery against the possible return associated with the guaranteed payment. 

\begin{figure}[t!]
    \centering
    \includegraphics[width=\linewidth]{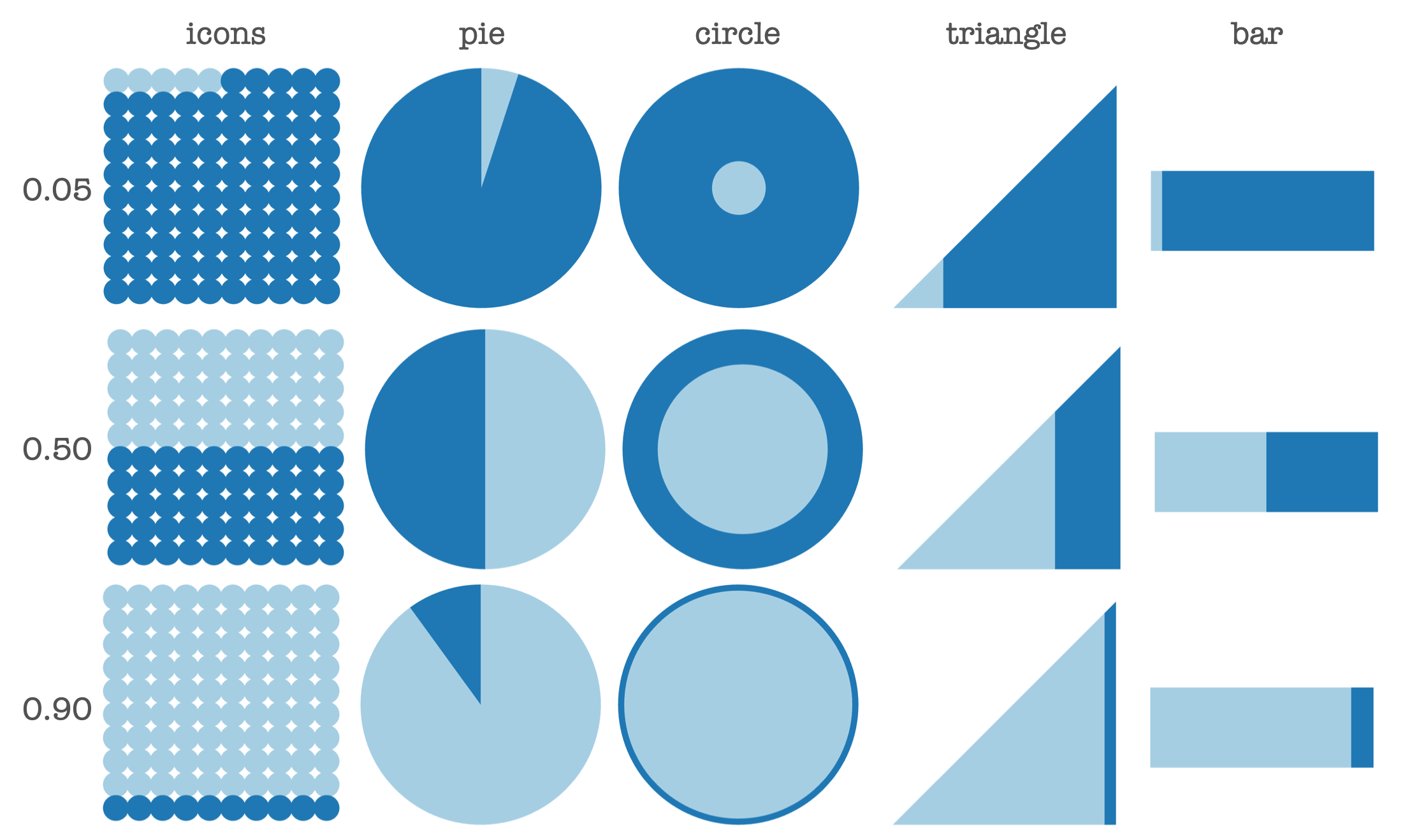}
    \caption{The visualization conditions used in the study. We randomly assigned each participants to one of the five visualization conditions or a text-only condition. Participants drew 25 hypothetical lotteries with each of the seven risk values $\{.05, .1, .25, .5, .75, .9, .95 \}$. Although we had no control over the display size, the default size of the visualizations were $200\times200$ pixels, and the colors were colorblind safe.}
    \label{fig:conditions}
\end{figure}

We report the findings of two crowd-sourced experiment using a gambling scenario with real monetary gains. Our study participants chose to either enter a draw or receive guaranteed monetary gains based on the probabilities shown by the visualization designs.
From Experiment 1, we analyzed the data from 406 participants\textquotesingle which resulted in 10,150 observations. 
Each participant saw 25 lottery scenarios and we evaluated the perceived risk of 5 different types of widely used visualizations (see Figure~\ref{fig:conditions}).
Experiment 2 focused on decision-making.
We analyzed 300 participants\textquotesingle decisions which resulted in 7500 observations. Using the same gambling platform, participants made 25 lottery choices to win a prize of up to \$10.
We then evaluated the quality of the choices for each of the five visualization conditions (plus a control group with no visualization).

By modeling risk perception and behavior for each visualization, we provide a guideline for designers to select the right chart to communicate information. We make the following contributions to the understanding of risk perception and decision-making with visualization:
    
\begin{itemize}
    \item We demonstrate that visualization design impacts the graphical perception and causes probability distortion, and provide a clear ranking of the tested designs based on their perceptual errors\\ $icons>pie>bar>triangle>circle$
    \item We show that, for every tested design, people consistently overestimated the size of small proportions and underestimate the size of larger portions. 
    \item  We show that charts influence risk behaviors. For low probabilities, we observed that visualization groups (e.g. triangle with 5\% and pie with 10\%) exhibited greater risk-seeking tendencies.
    \item We provide generalizable models for each visualization that can predict perception and decision-making for the first time
\end{itemize}

\section{Background}

\subsection{Graphical Perception}
 
\begin{figure}[b!]
    \centering
    \includegraphics[width=.4\textwidth]{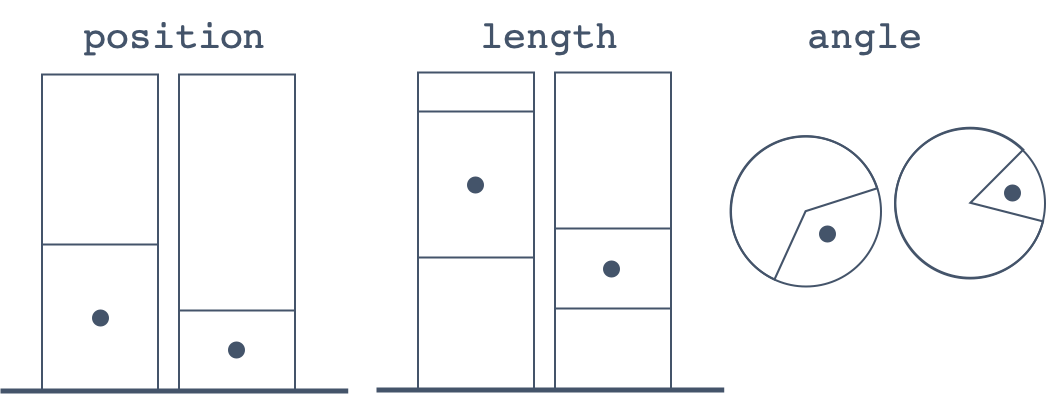}
    \caption{Charts used in prior works. Participants were asked to compare what proportion of the first probability is equal to the second. \protect\citeA{cleveland1984graphical,simkin1987information} }
    \label{fig:graphs}
\end{figure}

The act of decoding a data visualization is called \textit{graphical perception}~\cite{cleveland1984graphical}. There are a number of studies that investigated how visualization can lead to distortion in judgment. For instance, Cleveland and McGill provided a taxonomy of judgment that people make when decoding quantitative information from charts: position along a common scale, position along common scales, position along identical but non-aligned scales, length, angle, slope and area \cite{cleveland1984graphical,cleveland1986experiment,cleveland1987graphical}. They conducted a series of perceptual experiments where participants performed comparison judgment by estimating the percentage of a small division out of a larger one (see Figure~\ref{fig:graphs}). They found that people were best at decoding quantitative information when the data was encoded using the position of a visual mark and worst when using area.

Simkin and Hastie tested subjects' ability to make proportion (part-of-the-whole) judgment with different types of charts ~\cite{simkin1987information}. Participants estimated the percentage represented by the indicated portion of the chart. The results showed no measurable difference between the angle (pie) and position (bar) estimates, but angle and position were superior to length. Spence and Lewandowsky's study confirmed that there was no significance difference between angle and position visualizations~\cite{spence1991displaying}. 

Other researchers have also explored tasks such as judging correlations~\cite{harrison2014ranking,rensink2010perception,yang2019correlation} and perception in tabletop and multi-surface environments~\cite{wigdor2007perception}. Graphical perception has been extensively tested for various designs such as bar graph~\cite{zacks1998reading,skau2015evaluation,peck2013using}, pie chart~\cite{kosara2019impact,skau2016arcs,skau2016judgment,peck2013using}, bubble chart~\cite{heer2010crowdsourcing}, treemap~\cite{heer2010crowdsourcing}, scatterplot~\cite{lewandowsky1989discriminating,tremmel1995visual} and various representations of time series data~\cite{javed2010graphical}.

There are two primary findings to highlight from this diverse body of work: (1) visual encoding impacts graphical perception and (2) the nature of the task influences this interaction. Building on the prior research, we aim to investigate how visual encoding impacts risk perception, risk attitudes, and decision-making.

\subsection{Visualizing Risks for Decision-Making}

The existing research on visualizing risk and supporting decision-making is broad and spans many domains. Much of the prior work in the visualization community on supporting risk communication has focused on embedding data about uncertainty within the visualization itself~\cite{correll2014error,hullman2015hypothetical,kale2019hypothetical,kay2016ish,potter2011quantification}. Hullman et al. provides a comprehensive review of the existing work in this area~\cite{hullman2019pursuit}. There has also been a substantial body of work that investigates how visualization supports risk communication and medical decision-making (e.g., ~\cite{micallef2012assessing} and  ~\cite{ottley2016improving}). Typical evaluation methods utilize some measure of speed and accuracy. Very few have investigated the impact of visual design on decision-making.

Recent work by Kale et al. examined how people make decisions based on uncertainty visualizations~\cite{kale2019hypothetical}. In one study, participants played the role of a newspaper editor and saw either a bar chart with error bars displaying job statistics for a given year, or a \textit{hypothetical outcome plot}~\cite{hullman2015hypothetical} which uses animation to show the same data. Participants needed to decide between two headlines: "Latest Job Data Show \textbf{No Growth}" or "Latest Job Data Show \textbf{Growth Trend}". Their findings show that animation can effectively support decision-making under the tested circumstances. 

In the medical domain, Galesic et al. conducted an experiment where patients were shown the risk of a disease in a numerical or a visual format. They were asked to rate the perceived seriousness of the disease and the importance of screening on a scale of 1 to 15~\cite{galesic2009}. The results showed that participants who were shown the information in a numerical format rated the disease as more serious compared to the group who were shown icon arrays. Similarly, helpfulness of screening had a higher score for the numerical format. Galesic et al. states that visual aids help patients to make more informed medical decisions~\cite{galesic2009}. Ruiz et al. asked at risk patients to decide whether they would opt for screening based on visual risk information about the disease ~\cite{ruiz2013}. They confirmed the finding that people were more risk-averse when presented icon arrays. Other work by Nadav-Greenberg et al. investigated whether participants would issue a weather forecast when presented visual weather diagrams~\cite{greenberg2008}.


We situate our approach to investigate how visualization influences decision-making in \textit{Decision Theory}. We created a real-life gambling scenario and observe participants' lottery decisions to provide a ranking that can improve visualization selection to assist decision-making in a number of areas.

\subsection{Decision Theory}
Economists and psychologists have long studied how people make choices under risk by investigating prospects or gambling scenarios. A prospect is a contract:
\begin{equation}
    (x_1, p_1:...:x_n, p_n),
\end{equation}
\noindent
that yields $x_i$ with probability $p_i$, where $\sum_{i=1}^{n} p_i = 1$~\cite{kahneman2013prospect}. Prospects provide a simple model for understanding risky decisions. 

The classical method for evaluating a gamble is through assessing its expected value. The expected value of a prospect is the sum of the outcomes where the probabilities weight each value:
\begin{equation}
    ev = \sum_{i=1}^{n} p_ix_i
\end{equation}
\noindent
For example, consider the gambling scenario in Section~\ref{sec:introduction}, the expected value of option A is 500 ($.5 \times 1000 + .5 \times 0 $) and the expected value of option B is 450 ($1 \times 450$). A \textit{rational} decision-maker would then choose option A over option B. However, most people would choose the sure payment of \$450. This highlights the perhaps obvious conjecture that humans are not always rational.

One of the dominant theories of decision making, \textit{Expected Utility Theory} (EUT), has served for many years as both a model that describes economic behaviors~\cite{friedman1948utility} and a model of rational choice~\cite{keeney1993decisions}.  In particular, it states that people make choices based on their \textit{utility} - the psychological values of the outcomes. For instance, if a person prefers an apple over a banana, then it stands to reason that they would prefer a 5\% chance of winning an apple over a 5\% chance of winning a banana. Using EUT, we can assess the overall utility of a gamble by summing the utilities of the outcomes weighted by their probabilities. 
\begin{equation*}
    U = \sum_{i=1}^{n} p_i u(x_i)
\end{equation*}
\noindent
This model, however, still assumes that most humans are rational and consistent, and solely decide on prospects based on their utility~\cite{kahneman2013prospect}.
Still, EUT provides a tool for us to evaluated peoples' behavior when choosing among risky options and is the foundation for the other dominant economic theory, \textit{Prospect Theory}~\cite{kahneman2013prospect}.

A refinement of EUT that is used to describe risk perception and decision-making with risk empirically is known as Prospect Theory~\cite{kahneman2013prospect}. The theory essentially posits that people tend to underweight common or high-frequency events while over-weighting rare or low-frequency events. Typically, there is a probability weighting function $\pi$ such that 

\begin{gather*}
    \pi(p) > p,  \text{ when p is small, and} \\
    \pi(p) < p,  \text{ when p is large but not a certainty}
\end{gather*}

$\pi$ reflects the subjective desirability of a choice, which in practice replaces the stated probabilities with weighting factors $\pi(p)$. 
For example, researchers found that 72 out of 100 experiment participants favored the option of getting \$5000 with a probability of 0.001, a small probability event, over the prospect of getting \$5 for certain~\cite{kahneman2013prospect}. Both options have the same expected value, yet the majority of participants overestimated the probability associated with getting 5000. Furthermore, the prospect theory stipulates that such phenomenon has a two-fold impact on binary decision-making: (1) people tend to favor the option of getting a large gain with a small probability over getting a small gain with certainty, and (2) people tend to prefer a small loss with certainty over a large loss with tiny probability. Researchers see the former phenomenon as a risk-seeking behavior in the gain domain and the latter as a risk-aversion behavior in the loss domain. Lotteries can be expressed in terms of gains and losses (though it is uncommon) and therefore have related gain and loss domains.  For our purpose, we limit our scope to the gain domain. 

Relevant to the current work, Bruhin et al.~\cite{bruhin2010risk} conducted a series of large scale lottery studies and classified the distributions of behavioral types of different portions of the population based on how closely their behaviors are to that described by EUT and Prospect Theory. 
They analyzed the Relative Risk Premia, a descriptive metric of how risk-seeking or risk-averse a choice is~\cite{bruhin2010risk}. They showed that participants were risk-seeking for low-probability gains and risk-averse for high-probability gains. 

\begin{figure}[t]
    \centering
    \includegraphics[width=.45\textwidth]{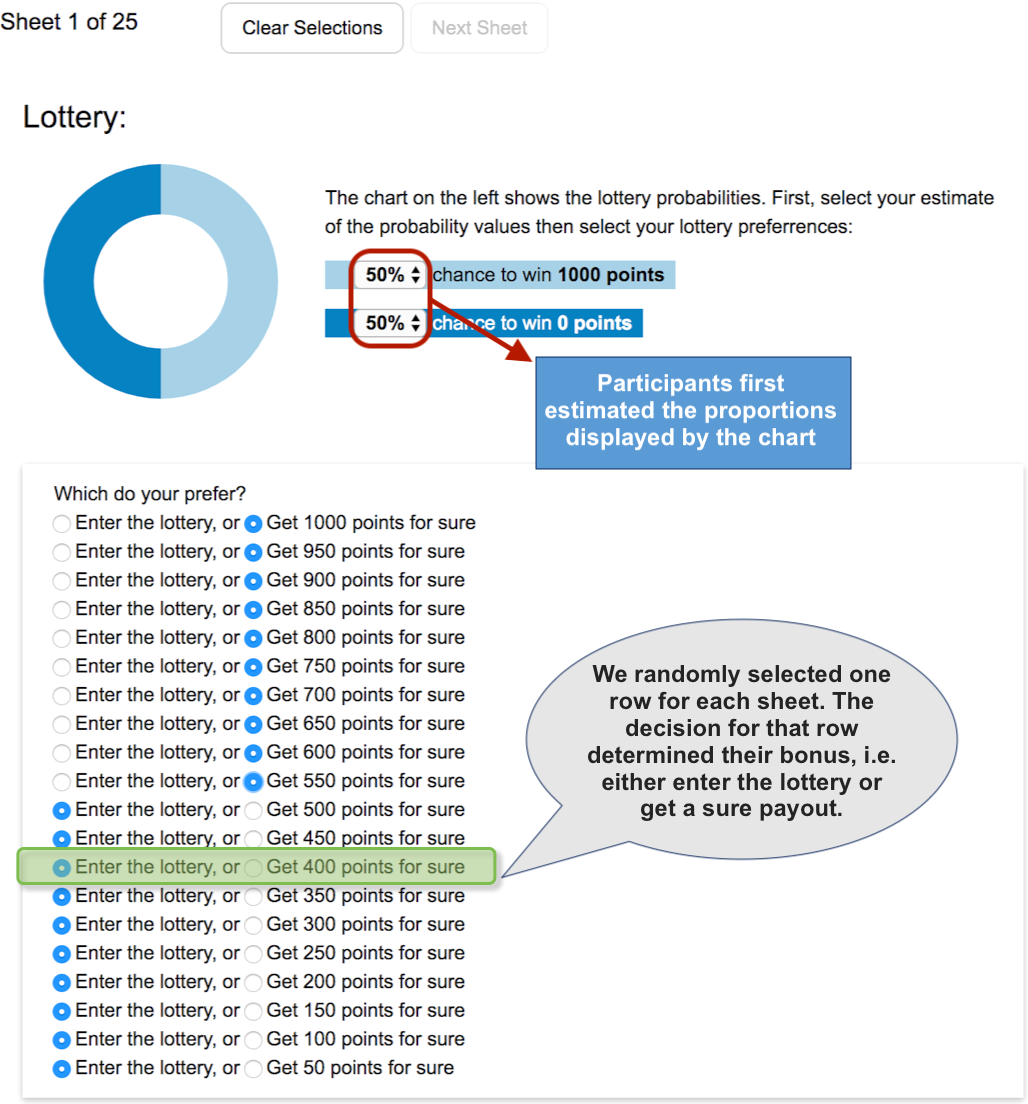}
    \caption{An example of the lottery decision sheet used in Experiment 1. Participants were asked to give their best guess for the proportion displayed by the chart. They only needed to select one value - the other one was automatically populated for them. Experiment II does not contain the probability estimation part but instead focuses on the lottery decisions}
    \label{fig:lotterysheet}
\end{figure}

\section{Research Questions}
This manuscript extends the prior research by investigating the complex relationship between visualization design, risk perception, and decision-making.
Despite the sizable body of work in graphical perception and risk visualization, we know very little about how encoding choices might influence real-world decisions. Prior work in psychology and economic theory provides a convenient foundation to investigate this relationship.   
In particular, we leverage a classical task for eliciting decision-making under risk, by observing actions as participants chose between entering a gamble and or receiving a guaranteed bonus payoff. We used five common designs to display seven lottery probability values that ranged from a 5\% to a 95\% chance of winning. We framed the following research questions to guide our investigation:
\begin{enumerate}
    \item How do visualization design and probability values impact risk perception and probability distortion?
    \item Does visual encoding impact risk attitudes and the choices people make?
\end{enumerate}

In the succeeding sections, we present the results of two large scale web-based experiments. Our first study investigates how visualization design influences probability distortion. Our second experiment evaluates how people make decisions from visualization.
Together, these experiments address the question of how users decode and make choices based on various visualization designs.

\section{Experiment 1 : Proportion Distortion}

Decision-making with visualization first involves decoding the information that is displayed. 
Thus, we begin our investigations by inspecting the effect of visualization on probability perception. Although Experiment I focuses on perceptual errors, we used the gambling scenario to maintain situational consistency across the two studies. We randomly assigned each participant to one of five visualization groups (pie, icon, triangle, bar and circle) and within each group, they observed seven probabilities. Participants completed twenty-five lottery sheets as demonstrated by Figure~\ref{fig:lotterysheet}.
They first estimated the lottery probabilities by entering their best guess of the values depicted by the visualization. Then they chose between entering the lottery or receiving a sure bonus payout.  

\subsection{Participants}
We recruited 406 participants from Amazon Mechanical Turk. There were 234 men and and self-reported ages ranged from 18 to 65 years ($\mu = 34.7;  \sigma = 9.49$). 73\% of our participants self-reported to have completed at least a college education.

Each participant had a HIT approval rate of 98\% with at least 100 approved HITs. We did not limit geographical location, but we required all subjects to be English-speaking between 18 and 65 years. We paid a base rate of \$1.00, plus a bonus of up to \$10.70 depending on the lottery outcomes. Bonus payments range from \$4.35 to \$9.30 ($\mu = 7.18 ; \sigma = .94$). Each participant worked at their own speed and the average completion time was approximately 29 minutes.  

\begin{table}[b!]
\caption{The prospects that were used in the study. $p_1$ denotes the probabilities ($p_2 = 1 - p_1$) and $x_1$ and $x_2$ are the outcomes. We used a point system for the outcomes where 1 point equaled \$0.01. Participants saw all combinations in a random order.}
\label{tab:prospects}
\begin{tabular}[.4\textwidth]{|llll|lllll|llll|}
\hline
$p_1$   & $x_1$  & $x_2$ &  &  & $p_1$   & $x_1$  & $x_2$ &  &  & $p_1$   & $x_1$ & $x_2$ \\ 
\hline
.05 & 20  & 0  &  &  & .25 & 50  & 20 &  &  & .75 & 50 & 20 \\
.05 & 40  & 10 &  &  & .50  & 10  & 0  &  &  & .90  & 10 & 0  \\
.05 & 50  & 20 &  &  & .50  & 20  & 10 &  &  & .90  & 20 & 10 \\
.05 & 150 & 50 &  &  & .50  & 40  & 10 &  &  & .90  & 50 & 0  \\
.10  & 10  & 0  &  &  & .50  & 50  & 0  &  &  & .95 & 20 & 0  \\
.10  & 20  & 10 &  &  & .50  & 50  & 20 &  &  & .95 & 40 & 10 \\
.10  & 50  & 0  &  &  & .50  & 150 & 0  &  &  & .95 & 50 & 20 \\
.25 & 20  & 10 &  &  & .75 & 20  & 0  &  &  &     &    &    \\
.25 & 40  & 10 &  &  & .75 & 40  & 10 &  &  &     &    &   \\
\hline
\end{tabular}
\end{table}

\begin{figure*}[t!]
    \centering
    \includegraphics[width=.8\textwidth]{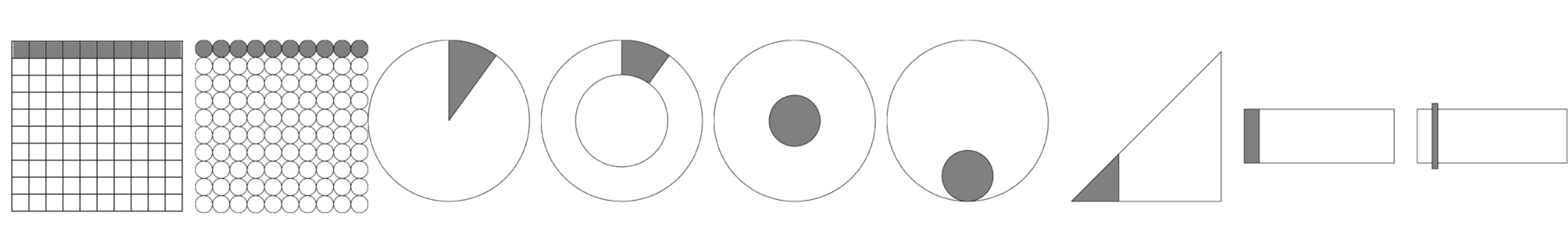}
    \caption{The initial designs that we considered for the study. After accounting for the nature of the charts (area, length, natural frequency), we selected designs 2,3,4,7 and 8.}
    \label{fig:initial-designs}
\end{figure*}

\subsection{Lottery Game}
Consistent with prior work from the economic decision-making domain~\cite{bruhin2010risk}, we presented participants with 25 two-outcome lotteries ($n=2$) that were choices between risky and certain gains. We used a points system for our payoff quantities where 1 point equaled \$0.01. The probabilities, $p_i$, were drawn from the set $P = \{.05, .1, .25, .5, .75, .9, .95 \}$ and the outcomes $x_1$ and $x_2$ ranged from 0 to 150 points (\$0 to \$1.50). Table~\ref{tab:prospects}, summarizes the probability and outcome combinations used in the study. 

Each lottery sheet comprised of the prospect to consider and a list of 20 equally-spaced outcomes that ranged from $x_1$ to $x_2$. Figure~\ref{fig:lotterysheet} shows an example of the lottery sheets. Participants first estimated the lottery probability values, then used the decision sheet to select their lottery preferences. For each row, they indicated whether they preferred to enter the lottery or receive a guaranteed payment. For the control condition (no visualization), participant saw the lottery prospects in text and simply selected their preferences.

At the end of the experiment, we randomly selected one row from each of the 25 decision sheets, and the participant's choice in that row determined the bonus. For example, let us assume that we randomly drew the row highlighted in green in figure~\ref{fig:lotterysheet}. Suppose that the participant indicated a preference for the guaranteed payment; her bonus for that sheet would be 400 points (\$4.00). Now, suppose instead, the participant opted to enter the lottery, we would simulate the lottery to determine her payment. The bonus was the sum of winnings for each sheet.

\subsection{Visualization Designs}
Displaying proportions is an everyday use of data visualization. 
For instance, visualizations are prominent in the medical community for communicating risks of diseases and screening test results. Similar proportion charts are used daily in press, business, and market surveys. 

There are many potential designs for representing probabilities. For inspiration, we reviewed common visualizations in the medical community for displaying risks. Figure~\ref{fig:initial-designs} shows the initial designs that were considered in for the study. We selected nine possible encodings of proportions and conducted a series of pilot studies to narrow the selections to unique designs. We rejected designs based on judgment similarity (e.g., the grid and icon array elicited similar counting judgment) or due to inconsistency with the design space. Informed by our pilot studies, we chose the five visualizations in figure~\ref{fig:conditions}: \textit{icons}, \textit{pie}, \textit{circle}, \textit{triangle}, and \textit{bar}.







We selected these visualization designs because they are widely used in different fields and represent area, length and proportion judgment~\cite{cleveland1984graphical}. \textit{Icons} are in the medical community to communicate risks of diseases and importance of screening ~\cite{galesic2009}. They represent natural frequencies (i.e., 1 out of 10 versus 10\%) and prior work has shown that they facilitate logical reasoning~\cite{gigerenzer1995improve} and accurately represent probabilities ~\cite{micallef2012assessing,ottley2015improving}. \textit{Pie} and donut charts are widely used in various fields such as information graphics and market surveys. 
Although there are many variants of the pie and donut charts~\cite{skau2016arcs,kosara2019impact}, we opted for the standard designs. 
The \textit{triangle}, \textit{circle}, and \textit{bar} are all common in the medical community~\cite{vizhealth}, and represent area and length judgment~\cite{cleveland1984graphical}.

\subsection{Procedure}
After selecting the task on Mechanical Turk, participants consented per [redacted for anonymity] IRB protocol. They read the instructions for the study and completed one trial round using the text-only example shown on page~\pageref{example-prospect}. They then saw a short tutorial that explained the selections and the bonus calculation. Specifically, we explained that participants should first enter their best guess for the visualized probabilities. They only needed to select one of the options. The system automatically completed the other. Similarly, for the lottery choices, they only needed to select one of the options where they switched either decision. The system automatically populated the remaining radio buttons.  To prevent potential biasing, we used a donut chart for the instructions which was not a visualization condition in the study. Each participant was randomly assigned a visualization condition drawn from the set \{\textit{icons}, \textit{pie}, \textit{circle}, \textit{triangle}, and \textit{bar}\} and completed 25 tasks where they entered their best guess for the probability values displayed which were drawn from the set $P = \{.05, .1, .25, .5, .75, .9, .95 \}$. The order of the sheets was counterbalanced to prevent ordering effects. 


\subsection{Measures}
The study resulted in 10,150 observations (406 participants each completed 25 lottery sheets). During the experiment, we recorded the participants' probability estimates and preference selections. The experiment included the following independent variables:
\begin{itemize}[noitemsep]
    \item \textbf{7 Probability Values:} $\{.05, .1, .25, .5, .75, .9, .95 \}$
    \item \textbf{5 Visualizations:} $\{icons, pie, circle, triangle, bar\}$
\end{itemize}

\noindent
Our dependent variables were:

\begin{itemize}[noitemsep]
    \item \textsc{\textbf{Exact: }} Binary true or false if the participant's probability estimate matches the true probability.
    \item \textsc{\textbf{Bias: }} The difference between the participant's probability estimate and the true probability value. 
    \[
        Bias = \hat{p} - p,
    \]
    where $\hat{p}$ is participant's estimate of the probability value and $p$ is the true probability. 
    \item \textsc{\textbf{Error: }} The absolute value of \textsc{Bias}
\end{itemize}

\subsection{Hypotheses}

Our hypotheses were:
\begin{itemize}[noitemsep]
    \item \textbf{H1a:}  \textit{Icons} group will yield the lowest \textsc{Error}. Participants in the icon group will be best at estimating the true probability values and will be least likely to make errors. 
    \item \textbf{H1b:} We will replicate the findings from existing body of work on graphical perception. Participants will be better at estimated probabilities with \textit{pie} than with \textit{circle}, \textit{triangle}, and \textit{bar}. The anchor points in pie (e.g.: 25\%, 50\%, and 75\%) will ease judgment~\cite{simkin1987information}. We also we hypothesize that estimates with \textit{circle} and \textit{triangle} will be worse overall, as area judgment is the most difficult graphical perception task~\cite{cleveland1984graphical}. 

\end{itemize}

\begin{figure}[t!]
    \centering
    \includegraphics[width=.46\textwidth]{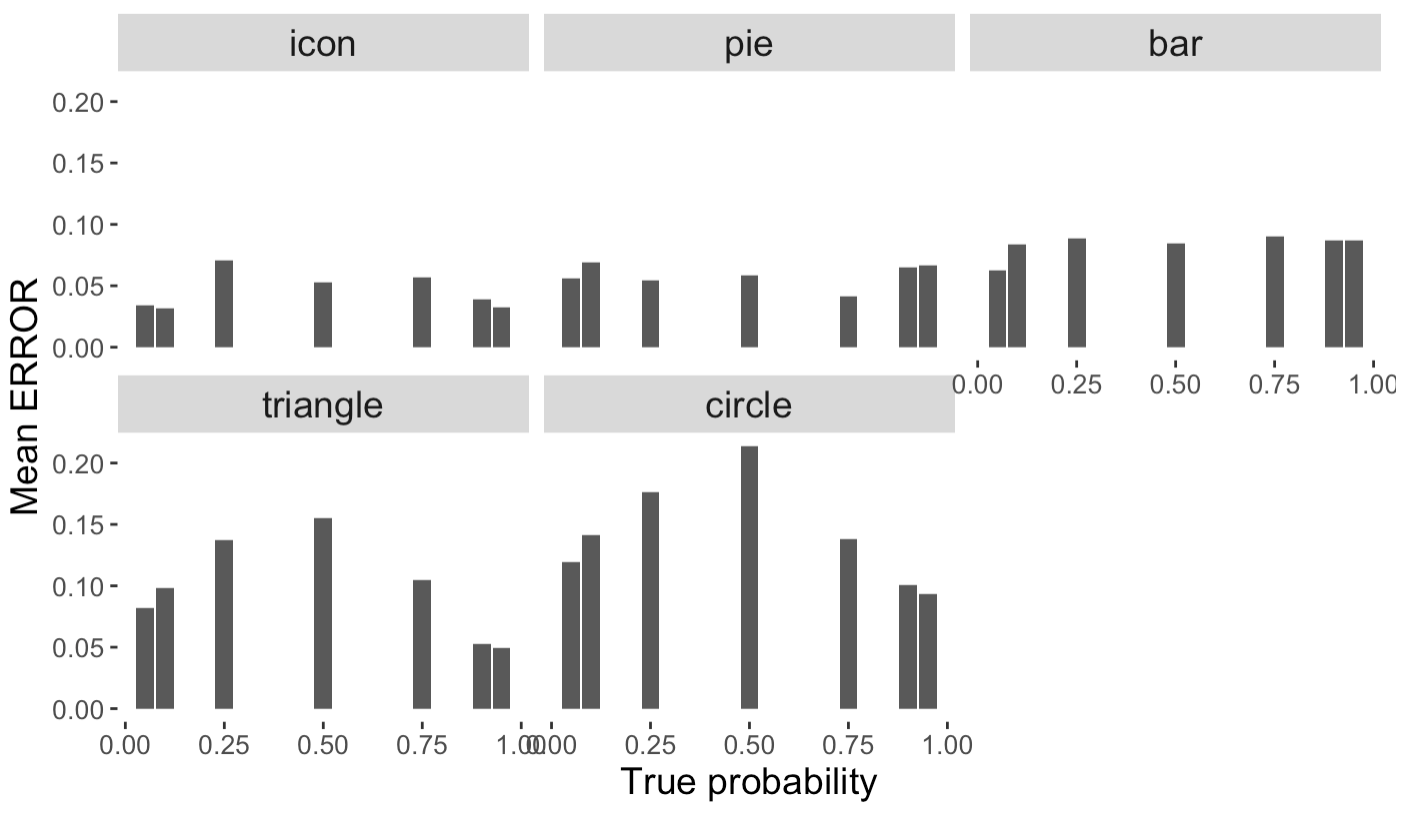}
    \caption{The mean error at each probability value for each visualization condition. We observe how probability values have an effect on probability distortion and how these variations vary across designs}
    \label{fig:error-distributions}
\end{figure}

\subsection{Results}

We begin our analysis by investigating \textsc{Exact} answers for the five different types of charts. We found that participants estimated the \textsc{Exact} probability 40\% of the time. Subjects in the \textit{icons} group produced \textsc{Exact} estimates 72.9\% of times, performing better than all other conditions. \textit{Pie}, \textit{bar}, \textit{triangle} and \textit{circle} yielded 59.8\%, 39.4\%, 13.7\%, and 10.6\% respectively. 

\subsubsection{Ranking by Error}
We then analyzed \textsc{Error} which captures the discrepancy between the participants' estimates and the exact proportions. We observed overall mean \textsc{Error} scores of 0.0463 ($\sigma = .12$), 0.0586 ($\sigma = .12$), 0.1497 ($\sigma = .14$), 0.1050 ($\sigma = .11$), and 0.0831 ($\sigma = .13$) for \textit{icons}, \textit{pie}, \textit{circle}, \textit{triangle}, and \textit{bar} respectively.
We conducted a comprehensive set of analyses to investigate how visualization and probability values impacted participants' \textsc{Error}.
Figure~\ref{fig:error-distributions} summarizes our findings.

A Kruskal-Wallis H non-parametric test revealed a statistically significant difference in the overall \textsc{Error} across conditions, $\chi^2(4, N=8475) = 1,412.61, p < .001 $. Pairwise Wilcoxon Mann-Whitney tests with a Bonferroni-adjusted alpha ($\alpha = 0.0023$) uncovered that differences in accuracy between all pairs of charts were statistically different with adjusted p-values less than $.001$, suggesting a strict ordering \textbf{\textit{icons}\textgreater\textit{pie}\textgreater\textit{bar}\textgreater\textit{triangle}\textgreater  \textit{circle}}.

The observed mean \textsc{Error} for across probabilities $.05, .1, .25, .5, .75, .9,$ and $.95$ were 0.0558 ($\sigma = .11$), 0.0671 ($\sigma = .12$), 0.0848 ($\sigma = .13$), 0.0931($\sigma = .14$) 0.0703($\sigma = .09$), 0.0558($\sigma = .11$) and 0.0531($\sigma = .11$) respectively (see Figure 6). We also observe that \textsc{Error} is the highest at $.5$ for conditions \textit{circle} and \textit{triangle}.

\begin{figure}[t!]
    \centering
    \includegraphics[width=.46\textwidth]{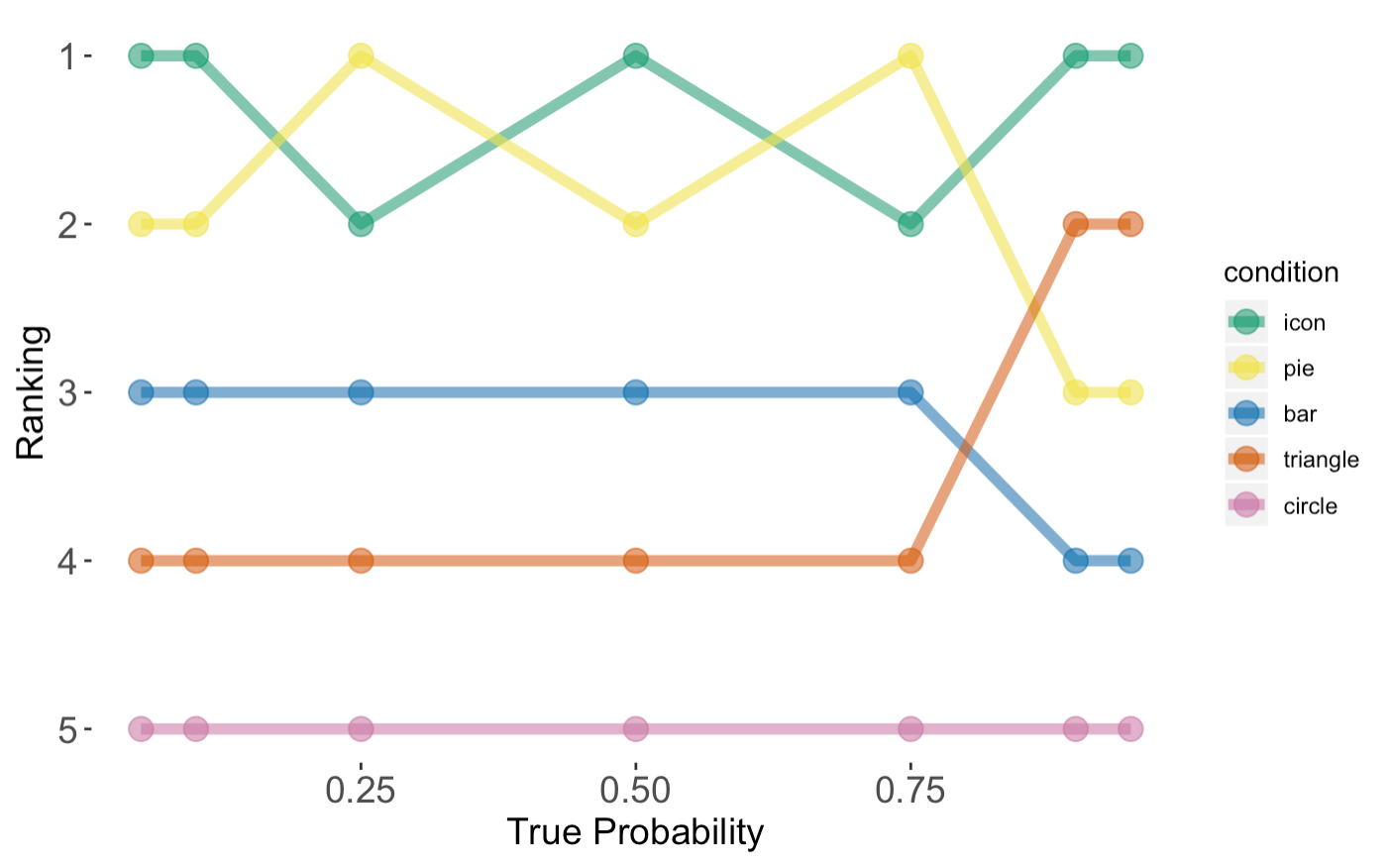}
    \caption{The ranking of visualization conditions based on mean error across probability values}
    \label{fig:ranking-error}
\end{figure}


A series of Kruskal-Wallis H non-parametric test confirmed that \textsc{Error} across the range of probability values were all significantly different with $p < .001 $. To provide a ranking of the visualization conditions at each probability value, we conducted Pairwise Wilcoxon Mann-Whitney tests with a Bonferroni-adjusted alpha to test for significant differences in \textsc{Error} among the charts at each probability value. Based on the results, we provided a ranking shown in Figure~\ref{fig:ranking-error}.

\subsubsection{Proportion Estimate Model Comparison}

\begin{figure}[b]
    \centering
    \includegraphics[width=.45\textwidth]{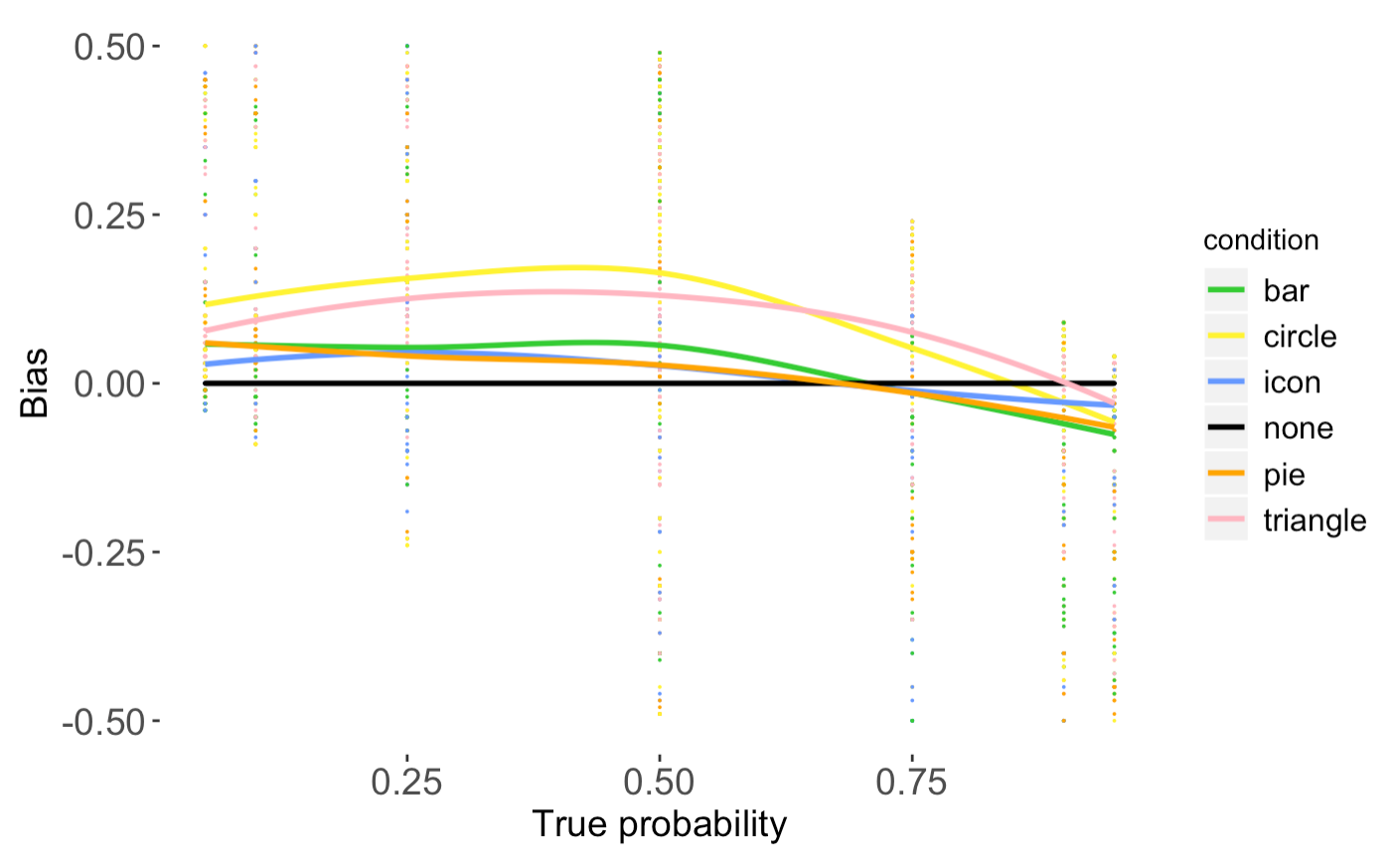}
    \caption{The bias across probability values for each design. We observe that \textsc{Bias} is positive across most probability values showing an overestimation in proportion estimate. For very high probabilities, proportion estimate is most often underestimated}
    \label{fig:bias}
\end{figure}

Our final analysis sought to examine how visualization design impacts probability distortion across different probability values.
We conducted a series of simple linear regressions~\footnote{Although we report only a linear regression, our analysis also examined a log-linear and a linear log odd model using individual observations. We determined, based on the AIC and deviance values, that the linear model was the best fit for every chart.} to examine whether it is possible to predict \textsc{Bias} given the true probability values ($p$) for each chart. The linear model take the form:
\[
    y_v = \beta_{v0} + \beta_{v1}x,
\]
where $y$ represents the outcome variable, and $x$ is the true probability. Table~\ref{tab:linear_model} summarizes the findings.
The results of the regressions indicate that the models explained 70\% to 85\% of the variance and all models were significant with $p<001$, providing evidence that $p$ predicted \textsc{Bias}.  Figure~\ref{fig:bias} shows the linear models for each visualization conditions. The final predictive models were:

\begin{itemize}[noitemsep]
    \item \textit{icons: $0.0520 + 0.9192p$}
    \item \textit{pie: $0.0766 + 0.8667p$}
    \item \textit{circle: $0.1878 + 0.7990p$ }
    \item \textit{triangle: $0.1353 + 0.8828p$}
    \item \textit{bar: $0.0923 + 0.8514p$}
\end{itemize}

\begin{table}[t!]
\caption{The rsquare, AIC, skewness, kurtosis and deviance values for the \textsc{ERROR} linear model for each design. These values were compared to a log linear and a linear log odds regression and showed a better fit.}
\label{tab:linear_model}
\resizebox{\linewidth}{!}{
\begin{tabular}[.4\textwidth]{|cccccc|}
\hline
$condition$  & $rsquare$ & $AIC$ &  $skewness$   & $kurtosis$  & $deviance$ \\ 
\hline

icon & 0.86  & -2326.88 & 0.79 & 10.02 & 24.39\\
pie & 0.83 & -2170.19 & 0.72 & 7.89 & 27.60  \\
circle & 0.70  & -1002.68 & -0.52 & 3.78 & 41.06\\
triangle  & 0.84 & -2167.39 & 0.34 & 6.07 & 27.09\\
bar & 0.78  & -1298.03  & 0.54 & 5.54 & 28.73	 \\
\hline
\end{tabular}
}
\end{table}

\subsection{Discussion}
At a high-level, the results from Experiment 1 provides three important contributions. First our findings serve as a \textit{re-examination} and \textit{replication} of prior work. Second, our analysis and experiment design provides meaningful extensions that are important to translate research findings into practical design guidelines. Third, we establish a baseline for subjective proportion estimates, allowing us to better isolate the impact that charts may have on decision-making.

\subsubsection{Icons are best.}
The results of our study reveal significant differences in risk perception across visualization representations. 
The notion that visual design biases risk perception is far from novel.
Differing from prior work, we tested proportion distortion with icon arrays, and showed that participants in the \textit{icons} group were most likely to estimate the exact proportions and made the smallest errors overall. 
Advocates of icon arrays believe that countable objects (natural frequencies) are more aligned with the way people think about proportions. 

Our results partially support findings of Cleveland and McGill~\cite{cleveland1984graphical,cleveland1986experiment,cleveland1987graphical}.
We observed that people in the \textit{triangle} and \textit{circle} groups were generally worst at estimating proportions,  corroborating that judgment with area proportioned charts were challenging for people in these groups.
Unlike Cleveland and McGill, we found that the \textit{pie} design performed reasonably well at minimizing proportion distortions. 
This is likely because the task was essentially a part-of-the-whole judgment~\cite{simkin1987information,spence1991displaying,skau2016arcs} and the 25\%, 50\% and 75\% anchor points provided useful visual cues.

\subsubsection{Overestimating the small and underestimating the large.}
An analysis of perception errors across the seven different probability values led to further discoveries. Unsurprisingly, people made significantly fewer errors with \textit{pie} when the probability was 25\%, 50\% or 75\%, with the lowest errors at 75\%. People made significantly more errors when the true probability was 50\% with the \textit{triangle} and \textit{circle} designs. 

Across all charts, we observed an overall tendency for the people to overestimate small percentages and underestimate large probabilities.
This phenomena of overestimating and underestimating can be found in many other environments~\cite{varey1990judgments,zhang2012}. 
For example, judging the frequency of letters in the English text~\cite{attneave1953psychological}, estimating word frequency~\cite{begg1974estimation}, estimating the proportion of white vs. black circles during a perceptual task~\cite{brooke1977error,varey1990judgments}, and as we will show in Experiment 2, estimating gambles~\cite{kahneman2013prospect}.
It is possible that proportion estimates with visualization design and these high-level judgment are governed by the same laws, and researchers have developed theories and models to account for this probability/frequency distortion effect~\cite{zhang2012}. 
However, future work is needed to investigate this phenomenon.

There are possibly many external components that come into play when observing probability distortion. It is likely influenced by factors such as the nature of the data, the environmental context or even individual characteristics. Future work is need to investigate the nuances of these effects. 
Still, our findings suggest that that icon array was the most effective visualization for representing percentages. Among all the charts we explored observed an ordering of \textbf{\textit{icons}\textgreater\textit{pie}\textgreater\textit{bar}\textgreater\textit{triangle}\textgreater  \textit{circle}}, based on people's accuracy in perceiving the true probabilities. 


\section{Experiment 2: Lotteries}
Our second experiment aims to uncover the impact of visualization design on decision-making. Consistent with Experiment 1, we used the gambling game to elicit choices. The lotteries in this experiment are structured similarly to the ones in Experiment 1. However, in this experiment we do not prompt participants to estimate the probabilities visualized as it is possible that this step could bias the lottery decisions that we observe.

\subsection{Participants}

We recruited 300 participants from Amazon Mechanical Turk. There were 193 men and and self-reported ages ranged from 19 to 65 years ($\mu= 34.0;\sigma=9.01$). $56\%$ of our participants self-reported to have completed at least a college education. The qualifications and bonus payments produces remained the same as Experiment 1.

\subsection{Procedure}
In a similar fashion to Experiment I, we presented participants with 25 two-outcome lotteries ($n=2$) that were choices between risky and certain gains. The probabilities, $p_i$, were drawn from the set $P = \{.05, .1, .25, .5, .75, .9, .95 \}$ and the outcomes $x_1$ and $x_2$ ranged from 0 to 150 points. Table~\ref{tab:prospects}, summarizes the probability and outcome combinations used in the study. 

Each participant was randomly assigned a visualization condition drawn from the set \{\textit{icons}, \textit{pie}, \textit{circle}, \textit{triangle}, \textit{bar} and \textit{none}\} where none is the text-only condition. 
We make two changes to lottery sheet design: (1) We removed the prompts for probability estimates; and (2) For the control condition (no visualization), participants saw the lottery prospects in text.

\begin{figure}[t]
    \centering
    \includegraphics[width=.46\textwidth]{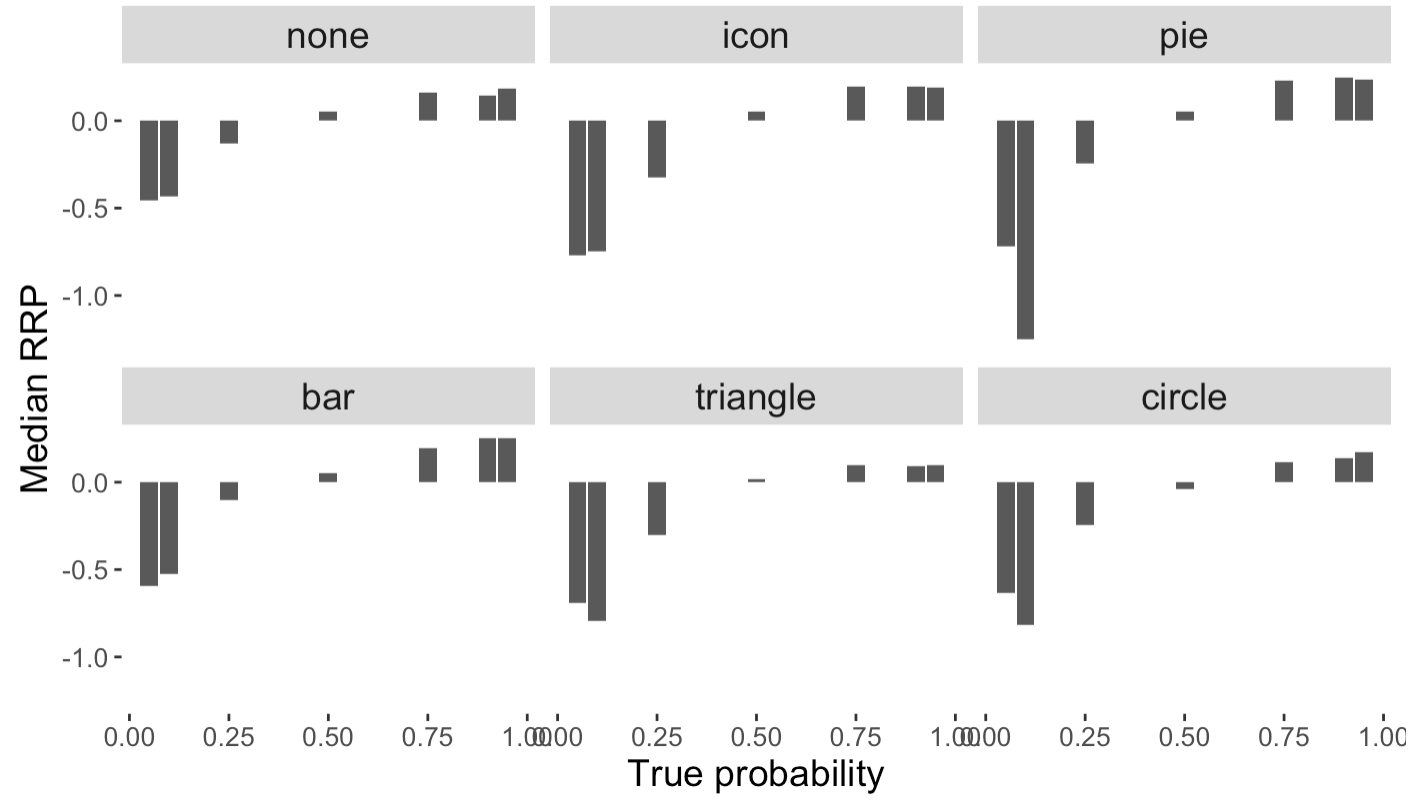}
    \caption{The median RRP values across probability values for each visualization condition. For all visualization conditions people were more risk seeking for low probabilities (RRP<0) and more risk averse for high probabilities (RRP<0). At 50\%, participants tended to be risk neutral   }
    \label{fig:median-RRP}
\end{figure}

\subsection{Measures}
The experiment included the following independent variables:
\begin{itemize}[noitemsep]
    \item \textbf{7 Probability Values:} $\{.05, .1, .25, .5, .75, .9, .95 \}$
    \item \textbf{6 Visualizations:} $\{none, icons, pie, circle, triangle, bar\}$
\end{itemize}

\noindent
To measure decision quality, our dependent variable was:

\begin{itemize}

    \item \textbf{RRP: } The Relative Risk Premia is used to evaluate the quality of the lottery decisions~\cite{bruhin2010risk} and can be seen as a measure of rationality. 
    
    \[
        RRP = (ev - ce )/ |ev|, 
    \]
    where ev denotes the expected value of the lottery outcome and ce is the certainty equivalent of the lottery. We can calculate the utility of a given prospect using the equation below:
    \[
        ev = \sum_{i=1}^{n} p_ix_i
    \]
    We calculate the lottery's certainty equivalent as the average of the smallest certain amount that the participant selected on the sheet and the subsequent certain amount on the sheet. For example, let us assume that figure~\ref{fig:lotterysheet} indicates a participant's selection. The certainty equivalent here is 525 ( $ce = (550 + 500) / 2$ ). \\
    
    $RRP > 0$ indicates risk aversion, $RRP < 0$ implies risk seeking behavior and $RRP = 0$ suggests risk neutrality. It is important to note that the RRP is independent of the perceived probability values. 
    
\end{itemize}

\subsection{Hypotheses}

Following our experiment design, we formed the following hypotheses:

Our hypotheses were:
\begin{itemize}[noitemsep]
    \item \textbf{H2a:}  The decision that we observe will follow Prospect Theory. In particular, we anticipate that participants will be risk-seeking for small probabilities ($RRP < 0$) and risk-averse for large probabilities ($RRP > 0$) gains~\cite{kahneman2013prospect}.
    \item \textbf{H2b:} We expect to observe $RRP_{icons} > RRP_{none}$. Previous research has shown that people are more risk averse when presented icon arrays compared to text ~\cite{ruiz2013}. Therefore, we anticipate that participants in the \textit{icons} group will be less risk-taking compared to the \textit{none} group.
    \item \textbf{H2c:} Based on the data from our Experiment 1, we hypothesize that perceptual errors with visualization design will influence behaviour. For example, we expect to observe that $RRP_{none} \neq RRP_{circle}$ and $RRP_{none} \neq RRP_{triangle}$. 

\end{itemize}






\begin{figure}[t]
    \centering
    \includegraphics[width=.46\textwidth]{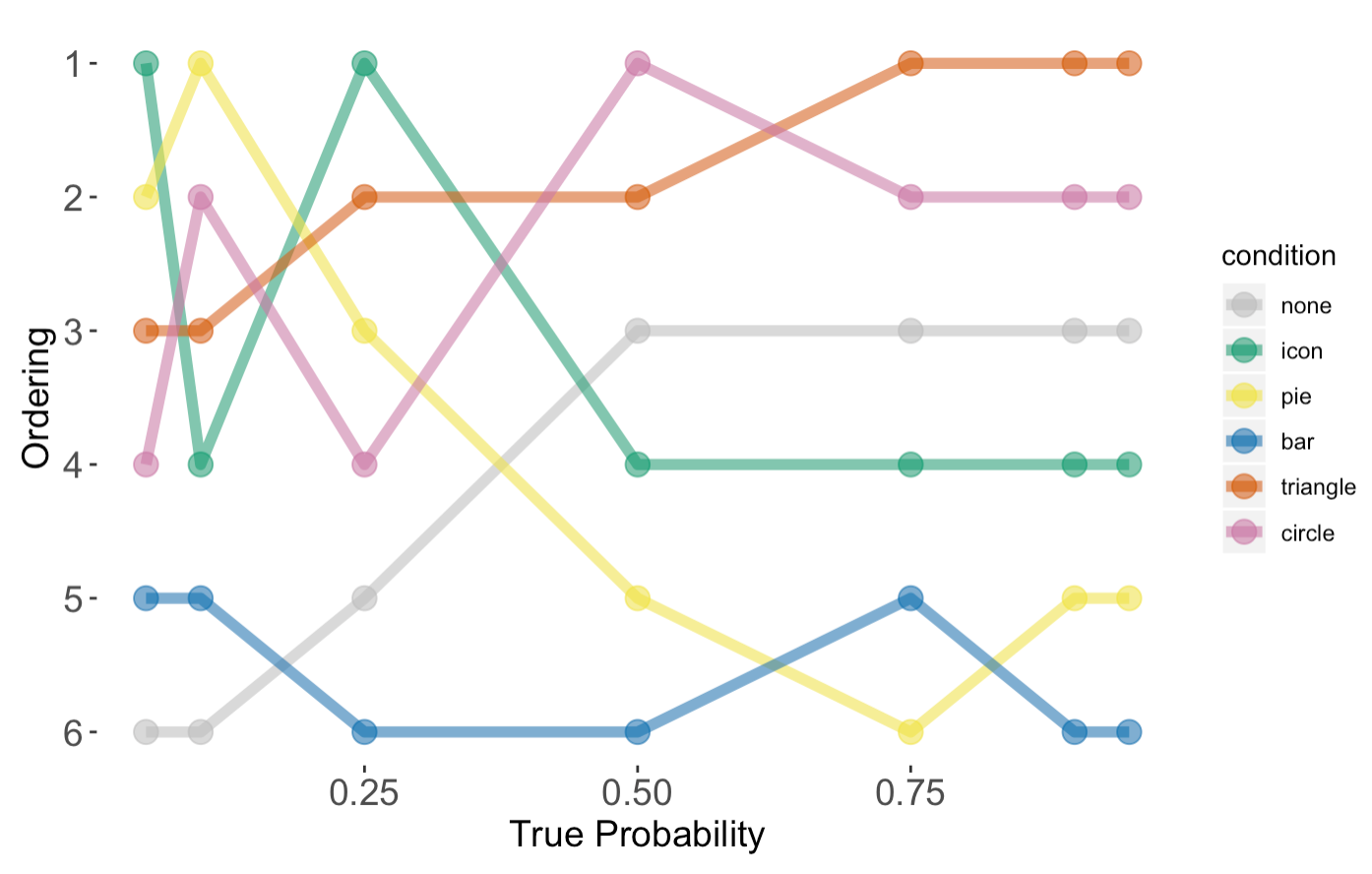}
    \caption{The ordering of visualization conditions based on median RRP across probability values.}
    \label{fig:RRP-ranking}
\end{figure}
    \label{fig:RRP-ranking}

\subsection{Results}

We conducted a fine-grained analysis of users' decisions across visualization conditions and probability values, using the Relative Risk Premia (RRP) to assess risk attitudes and decision-making behavior.

We observed overall median \textsc{RRP} values of 0.0167, 0.0000, 0.0155, 0.0273, -0.6625, and -0.0500 for \textit{none} \textit{icons}, \textit{pie}, \textit{bar},\textit{triangle} and \textit{circle} respectively. Figure~\ref{fig:median-RRP} shows the median RRP values across different probability values for each visualization design. We can observe that for all designs, users were risk seeking for low probabilities (RRP < 0) and risk averse for high probabilities (RRP > 0 ).



A Kruskal-Wallis H non-parametric test revealed statistically significant differences in \textsc{RRP} across conditions, $\chi^2(4, N=8475) = 3408.1, p = 2.2e-16 $. To provide an ordering of the visualization conditions based on the observed corresponding risk behavior at each probability value, we first ran seperate  Kruskal-Wallis H non-parametric test for each probalitity value then pairwise Wilcoxon Mann-Whitney tests with a Bonferroni-adjusted alphas. Based on the results, we provided a ranking shown in Figure~\ref{fig:RRP-ranking}. 



\subsection{Model Comparison}

Our analysis examined a linear and a logarithmic regression using individual observations. We determined based on AIC and deviance that the logarithmic model was the best fit for every chart. The logarithmic model takes the form:
\[
    y_v = \beta_{v0} + \beta_{v1}\ln(x),
\]
where $y$ represents the outcome variable and $x$ is the true probability. Table~\ref{tab:logarithm_model} summarizes the findings. All models were significant with $p<001$, providing evidence that $p$ predicted \textsc{RRP}. Figure 8 shows the logarithmic regression capturing the behavior of RRP across probability values. The final predictive models were:

\begin{itemize}[noitemsep]
    \item \textit{icons: $0.5296 + 1.0313\ln(p)$}
    \item \textit{pie: $0.5243 + 0.9653\ln(p)$}
    \item \textit{circle: $0.4411 + 0.9029\ln(p)$ }
    \item \textit{triangle: $0.1353 + 0.8828\ln(p)$}
    \item \textit{bar: $0.5281 + 0.8236\ln(p)$}
\end{itemize}



\begin{figure}[t]
    \centering
    \includegraphics[width=.46\textwidth]{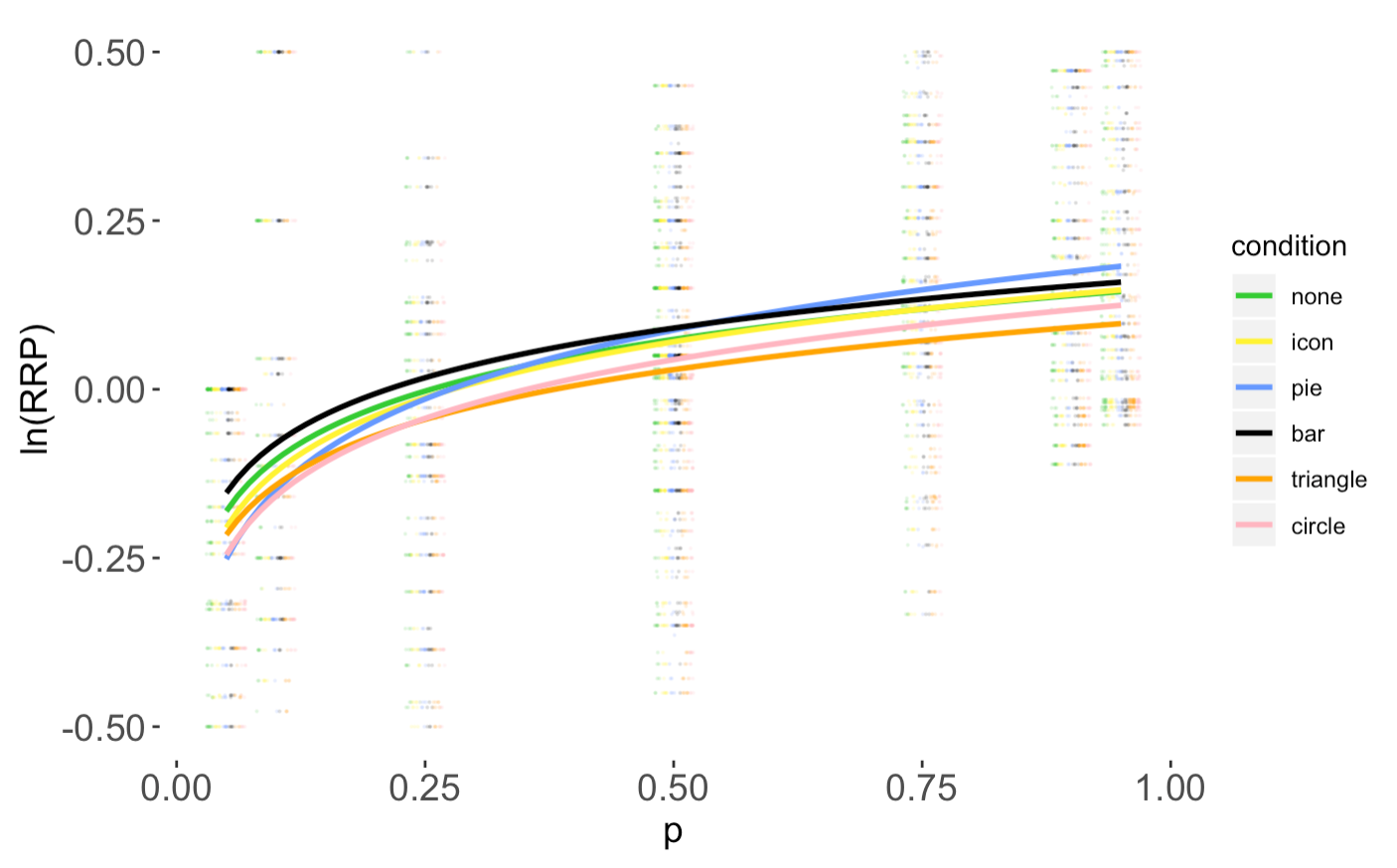}
    \caption{The logarithmic regression model showing the behavior of RRP across probability values for each visualization condition. This model has a better goodness-of-fit compare to a linear model.   }
    \label{fig:RRP}
\end{figure}


\subsection{Discussion}
The overall patterns in decision-making are clear.
Across all conditions, participants were risk-seeking with low-probability values and risk-averse with high probabilities. This replicates prior findings in  the economics domain~\cite{bruhin2010risk} and confirms Prospect Theory~\cite{kahneman2013prospect}.

A closer look at the \textsc{RRP} values across the different visualization conditions revealed a more nuanced view of visualization-mediated decisions. 
Our findings provide suggestive evidence that visualization design can indeed impact the decisions that people make. From Experiment 1, we observed that the \textit{circle}
and \textit{triangle} designs were most likely of the conditions to distort the true probability values. Similarly, the results of Experiment 2 showed that the lottery decisions made by the \textit{circle} and \textit{triangle} groups deviated significantly for the control condition (no visualization).

\begin{table}[t!]
\caption{The rsquare, AIC, skewness, kurtosis and deviance values for the RRP logarithmic regression model for each design.  These values were compared to a simple linear regression and showed a better goodness-of-fit based on AIC and deviance values.}
\label{tab:logarithm_model}
\resizebox{\linewidth}{!}{
\begin{tabular}[.4\textwidth]{|cccccc|}
\hline
$condition$  & $rsquare$ & $AIC$ &  $skewness$   & $kurtosis$  & $deviance$ \\ 
\hline

icon & 0.20  & 6996.28 & 4.52 & 30.24 & 7396.81\\
pie & 0.25 & 4869.07 & -3.79 & 24.57 & 3799.68  \\
circle & 0.23  & 3960.95 & -4.05 & 27.16 & 3074.05\\
triangle  & 0.20 & 5476.71 & -4.35 & 29.59 & 5117.81\\
bar & 0.20  & 4841.80  & -5.03 & 39.10 & 3716.02	 \\
\hline
\end{tabular}
}
\end{table}

Our results only partially confirms the findings of prior work that indicated that icon arrays may lead to more risk-averse behavior~\cite{ruiz2013}. We found that participants in the \textit{icons} group were markedly more risk averse than the \textit{none} group when $p = .05$, but we observed no significant difference overall. This is likely due to the contextual differences in the experiment designs and measures of decision making. In prior studies participants decided whether or not they would opt for screening based on visual risk information about the disease~\cite{galesic2009}. Future work is need to investigate the effect scenario on risk taking behavior.

Although RRP provides a convenient measure for us to summarize and classify risk taking behavior, we hesitate to draw conclusions about the \textit{quality} of the decisions that we observed. The desirability of the risk-seeking or risk-averse behavior is context dependent. Still, risk neutrality is an objective measure of rational decision-making, and using RRP allows us quantify and model departures from risk neutrality.

It should be noted that the visualizations chosen for this study are all commonly used in data visualization tasks across numerous disciplines and applications. Gambling, as presented it in this experiment, is a basic judgment task similar to the ones that people typically perform in everyday situations. For example, similar charts are often used to communicate medical risks and to support patients' decision making. Consequently, discrepancies in probability perception and decision-making across visualization designs and probability values are substantial findings that have a direct impact on the design and evaluation of risk visualization.


\section{Conclusion}
We used a crowed-sourced real-life lottery game to assess the effect of five visualization designs on probability distortion, risk-perception and decision-making. Our findings showed significant differences in perceptual errors across the visualizations conditions, showing that visualization mediates graphical perception which is the main cause of probability distortion. Our results also demonstrate that across all conditions, people tend to overestimate small probabilities and underestimate large probability values. Using the perceptual errors, we rank the representations based on subjects' accuracy in decoding quantitative information for each visualization condition across all probability values, and on average: \textbf{\textit{icons}\textgreater\textit{pie}\textgreater\textit{bar}\textgreater\textit{triangle}\textgreater \textit{circle}}. 

An analysis of gambling decisions found some consistent behavioral patterns across all visualization
conditions. We found that participants were risk-seeking with low-probabilities and risk averse with high probabilities. Furthermore, our results show a significant effect of visualization on risk attitudes. For instance, subjects in the circle and triangle groups exhibited greater risk-seeking behavior than participants in the text-only condition. Similar to the perception analysis, we rank the visualization conditions based on how they mediate risk and overall: \textbf{\textit{none} >= \textit{icons} >= \textit{bar} >= \textit{circle} >= \textit{pie} >= \textit{triangle}}. We showed that this ranking changes across probability values, where some conditions are more consistent while others vary in risk-behavior across probability values.  We believe that these results can have practical impacts on the design and evaluation of data visualizations to assist decision-making process with risk.

\balance{}

\bibliographystyle{SIGCHI-Reference-Format}
\bibliography{lets-gamble}

\end{document}